\documentclass{epl}

\usepackage{epsfig}

\usepackage{amsmath}
\usepackage{amssymb}

\usepackage[pdftitle={Persistence length of semiflexible polymers},
            pdfauthor={Petra Gutjahr,  Reinhard Lipowsky, Jan Kierfeld},
            pdfkeywords={Persistence length of semiflexible polymers}
            ]{hyperref}

\newcommand{\bt}{{\bf t}}
\newcommand{\D}{\displaystyle}

\title{Persistence length of semiflexible polymers\\ 
       and  bending rigidity renormalization}
\shorttitle{Persistence length and $\kappa$-renormalization}

\author{Petra Gutjahr, Reinhard Lipowsky and Jan Kierfeld}
\shortauthor{P. Gutjahr {\it et al.}}
\institute{Max Planck Institute of Colloids and Interfaces, Science
  Park Golm, 14424 Potsdam, Germany}


\begin{document}

\maketitle

\begin{abstract}
The persistence length of semiflexible polymers
and one-dimensional fluid membranes   is obtained 
from the renormalization of their bending rigidity.
The renormalized bending rigidity is calculated
 using an exact real-space
functional renormalization group transformation based on a 
mapping to the one-dimensional Heisenberg model. 
The renormalized bending rigidity vanishes exponentially
at large length scales and its 
 asymptotic behaviour is used to define  the persistence length.
For semiflexible polymers, our result agrees with the  persistence length 
obtained using the asymptotic behaviour 
of tangent correlation functions. 
Our definition 
differs from the one
 commonly used for  fluid membranes, which is based 
on a perturbative  renormalization of the  bending rigidity. 
\end{abstract}

\section{Introduction}

Thermal fluctuations of two-dimensional fluid membranes and 
one-dimensional semiflexible polymers or filaments are governed
by their bending energy and can be characterized using the concept
of a {\em persistence length} $L_p$. 
In the absence of thermal fluctuations at zero temperature, 
fluid membranes  are planar and filaments are  straight because of
their bending rigidity. 
Sufficiently large and thermally fluctuating  membranes  
or  filaments  lose  their planar or straight conformation. 
Only subsystems of size  $L\ll L_p$ appear rigid and 
maintain an average  planar or straight conformation with a preferred
normal or tangent direction, respectively.  
Larger membranes or filaments of sizes  $L \gg L_p$, on the other hand,
appear flexible. 
In the ``semiflexible'' regime  
for which $L$ is smaller than or comparable
with $L_p$, 
 the statistical mechanics
is governed  by the competition of the thermal energy $T$ and 
the bending rigidity $\kappa$. 
Experimental values for  the persistence length of one-dimensional 
biological filaments vary from 50 nm for double-stranded DNA~\cite{taylor},
10 $\mu$m for actin filaments~\cite{gittes,kaes} up to the mm-range for 
microtubules~\cite{gittes}.
The persistence lengths of two-dimensional fluid membranes composed of
lipid bilayers are typically 
much larger than experimental length scales.

For semiflexible polymers with one internal dimension, 
$L_p$ is usually defined by the characteristic 
length scale for  the exponential decay  of the two-point 
correlation function between unit tangent vectors $\bt$ along 
the polymer. 
A continuous model for an inextensible semiflexible polymer of 
contour length $L$ is the worm-like chain (WLC) model~\cite{kratky}. 
In the WLC model the polymer is parametrized by its arc length $s$
($0<s<L$) and the polymer contour is completely determined by the 
field $\bt(s)$ of unit tangent vectors. 
The Hamiltonian is given by the bending energy 
\begin{equation}
{\cal H}  \{ \bt(s) \} = 
\frac{\kappa_0}{2}\int_0^L ds \, (\partial_s {\bf t})^2, 
 ~~~\mbox{with}~ {\bf t}^2(s)=1.
\label{continuous}
\end{equation} 
where $\kappa_0$ is the (unrenormalized) bending rigidity of
the model.
For a WLC embedded in $d$ spatial dimensions, the tangent correlation 
function is found to be \cite{LL}
\begin{equation}
\langle {\bf t} (s) \cdot {\bf t} (s') \rangle = e^{-|s-s'|/L_p}, 
~~~\mbox{with}~ 
  L_p =  \frac{2}{d-1}\frac{\kappa_0}{T}
.
\label{tangent}
\end{equation}
where $T$ is the temperature in energy units. 

For fluid membranes with two internal dimensions, the analogous
quantity is the correlation function of normal vectors.
An approximate result has been given in Ref.\ \cite{degennes}, 
but a rigorous treatment is missing due to the more involved 
differential geometry.
Surfaces cannot be fully determined by specifying an arbitrary 
set of normal vectors, but have to fulfill additional compatibility
conditions in terms of the metric and curvature tensors, the equations of 
Gauss, Mainardi and Codazzi, which ensure their continuity. 
Implementations of these constraints lead to a considerably more
complicated  field theory than (\ref{continuous}) describing 
a two-dimensional  fluid membrane in terms of its normal
and tangent vector fields \cite{CG05}. 

For  fluid membranes, 
an alternative definition of the persistence length 
$L_p$ has been given, which is linked 
to the effect of $\kappa$-renormalization.
The mode coupling between thermal shape fluctuations of 
different wave lengths modifies the large scale bending behavior, 
which can be described by an effective or {\em renormalized} 
bending rigidity $\kappa$. 
The renormalized $\kappa$  has been calculated using different 
perturbative  renormalization group (RG) approaches
\cite{helfrich,peliti,foerster,kleinert,helfrich98,pinnowhelfrich}. 
The results are still controversial:
Several authors \cite{helfrich,peliti,foerster,kleinert} find a 
thermal softening of the membrane with increasing length scales, 
but differing prefactors,
whereas Pinnow and Helfrich \cite{pinnowhelfrich} 
 obtained the opposite result. 
Furthermore, different definitions of the persistence
length are considered in these approaches: In Refs.\
\cite{peliti,foerster,kleinert},  $L_p$ is identified 
with the length scale, where the renormalized bending
rigidity $\kappa$ vanishes, while Helfrich and Pinnow defined
$L_p$ via the averaged absorbed area \cite{helfrich,pinnowhelfrich}.

In this Letter, we obtain an exact real-space RG 
 scheme for the bending rigidity of 
 a semiflexible polymer or a one-dimensional fluid membrane, 
which allows  us to  define the persistence length as the 
characteristic decay length of the renormalized bending rigidity. 

In principle, a perturbative result for the effective
$\kappa$ can be deduced from the RG analysis of
the one-dimensional nonlinear $\sigma$-model, 
which is equivalent to the WLC Hamiltonian (\ref{continuous}).
After a Wilson-type momentum-shell RG analysis, one obtains
the effective rigidity
(see, e.g., Refs.\ \cite{qft})
\begin{equation}
\frac{\kappa(\Lambda)}{T} = \frac{\kappa_0}{T} 
\left[ 1- \frac{T}{\kappa_0} \frac{d-2}{\pi}
  \Big\{\frac{1}{\Lambda}-\frac{1}{\Lambda_0}\Big\}
+ {\cal O}(T^2/\kappa_0^2) \right]
,
\label{sigma_model}
\end{equation}
which depends on the momentum $\Lambda$.
The parameter $\kappa_0= \kappa(\Lambda_0)$ is the `bare' 
coupling taken at the high momentum cut-off $\Lambda_0= {\pi}/{b_0}$, 
which is given by a 'lattice spacing' or bond length $b_0$. 
Using also $\Lambda = {\pi}/{\ell}$
we obtain the renormalized  $\kappa = \kappa(\ell)$
as a function of the length scale $\ell$.
Following the procedure previously used for membranes,
the persistence length can be defined via
\begin{equation}
\kappa(L_p) \equiv 0 ~~~\mbox{and, thus,}~~
L_p \simeq \frac{\pi^2}{d-2}\frac{\kappa_0}{T} 
.
\label{persistence_sigma}
\end{equation}
For the case of the polymer in the plane, the Hamiltonian
simplifies to a free or Gaussian field 
theory such that $\kappa=\kappa_0$ is unrenormalized 
to all orders in $\kappa_0/T$ and, thus, $L_p$ as defined via 
$\kappa(L_p) \equiv 0$ would become infinitely large.

A similar perturbative momentum-shell RG
procedure is possible in the so-called Monge parametrization of 
a weakly bent semiflexible polymer, analogous to the RG analysis 
for two-dimensional membranes \cite{peliti}. 
Then the polymer is parametrized 
by its projected length $x$ with $0<x<L_x$, where $L_x$ is the 
fixed projected length of the semiflexible polymer while its contour
length becomes a fluctuating quantity. 
The renormalized $\kappa=\kappa(\ell_x)$ becomes a function of the 
projected length scale $\ell_x$, which complicates
a comparison with the result (\ref{sigma_model}), which was derived in
an ensemble of fixed contour length.
Using the analogous criterion $\kappa(L_p)=0$ we obtain 
$L_p \simeq 2\pi^2\kappa_0/(3d-1)T$
within the Monge parametrization. 

A comparison of the RG results 
from the non-linear $\sigma$-model, see eq.\
(\ref{persistence_sigma}),
and the one obtained in the Monge parametrization with
eq.\ (\ref{tangent}) shows that the RG results for the 
persistence length $L_p$ are not compatible with the definition
using the tangent correlation function. 
This raises the general question which of the definitions should 
be preferred.

In this work we concentrate on a discrete description for
semiflexible polymers, which is equivalent to the one-dimensional
classical Heisenberg model. The advantage of this model is, that the
$\kappa$-renormalization as well as the tangent correlation function
are exactly computable in arbitrary dimensions $d$. Consequently a
direct comparison of the persistence length determined via
$\kappa$-renormalization and via the tangent correlation function is
possible. We introduce this model in the next section. The
$\kappa$-renormalization is carried out in a similar fashion as is
commonly used for Ising-like spin systems. In contrast to the
nonlinear $\sigma$-model, we find nontrivial results for
$\kappa(\ell)$ both in two and in three dimensions. As expected for an exact
result, $\kappa(\ell)$ is always positive and approaches zero only
asymptotically. Finally, we analyze the large scale behavior of $\kappa(\ell)$ 
leading to a power series of exponentials with the same decay length
as obtained for the tangent-tangent correlations. We define this length
scale to be the persistence length of the polymer.

\section{Theoretical model}

A discretization of the WLC Hamiltonian (\ref{continuous})
should preserve its local inextensibility. In addition, we want to 
use a discretized Hamiltonian  which is 
 locally  invariant with respect to  full rotations 
of single  tangents ${\bf t}_i$ -- in addition to the global
 rotational symmetry of the polymer as a whole. 
A suitable discrete model is an
inextensible semiflexible chain model 
as given by \cite{KNSL04}
\begin{equation} 
{\cal H} \{ {\bf t}_i \}   =  \frac{\kappa_0}{b_0} \sideset{}{_{i=1}^M}\sum
(1-{\bf t}_i \cdot {\bf t}_{i-1}),  
~~~\mbox{with}~ {\bf t}_i^2=1
,
\label{discrete}
\end{equation}
with $M$ bonds or chain segments of fixed length $b_0$.
The semiflexible chain model 
is equivalent to the one-dimensional classical Heisenberg model
(except for the first term, which represents a constant energy term)
describing a one-dimensional chain of classical spins.

The partition sum reads
\begin{equation}
{\cal Z}_M =  \left(\prod_{j=0}^{M}\int d{\bf t}_j\right)
   \exp[-{\cal H} \{ {\bf t}_j \} /T]
=  \left(\prod_{j=0}^{M}\int d{\bf t}_j\right)
    \sideset{}{_{i=1}^M}\prod T_{i,i-1} \;,
\end{equation}
where we have introduced the transfer matrix 
\begin{equation}
 T_{i,i-1} =  \exp\left[ -K_0 (1- {\bf t}_i \cdot {\bf t}_{i-1})\right]
, ~~~\mbox{with}~ K_0\equiv \kappa_0/{b_0 T}
.
\label{Tij}
\end{equation} 
We can parametrize the scalar product of unit tangent vectors using
the azimuthal angle difference $\Delta \theta_{i,i-1}$ 
as ${\bf t}_i \cdot {\bf t}_{i-1} = \cos(\Delta\theta_{i,i-1})$. 
Then the transfer matrix can be expanded as 
\begin{subequations} \label{expansions}
\begin{equation}
\begin{array}{ll}
\D T_{i,i-1} =  \sum_{m=- \infty}^{\infty}
   \lambda_m^{(0)}  e^{i m \Delta\theta_{i,i-1}},&
\D \lambda_m^{(0)} (K_0) =  e^{-K_0} I_m(K_0)
\end{array}
\end{equation}
in two dimensions and
\begin{equation}
\begin{array}{ll}
\D T_{i,i-1} =  \sum_{l=0}^{\infty}
  (2l+1) \lambda_l^{(0)}  P_l(\cos \Delta\theta_{i,i-1}),&
\D  \lambda_l^{(0)}(K_0) 
  = \sqrt{\frac{\pi}{2K_0}} e^{-K_0} I_{l+1/2}(K_0)
\end{array}
\end{equation}
\end{subequations}
in three dimensions,
where $I_k(x)$ denotes the modified Bessel function of the
first kind and $P_l(x)$ the Legendre polynomials \cite{AS}.
In the following, the sums $\sum_{m= -\infty}^{\infty}$ for $d=2$
and $\sum_{l=0}^{\infty} (2l+1)$  for $d=3$ are 
abbreviated by $\sum^{(d)}_n$.

For simplicity, we restricted our analysis to $d=2$  and $d=3$  spatial 
dimensions,
but our results can easily be generalized to arbitrary dimensions $d$: 
The transfer matrix is then expanded in Gegenbauer polynomials and the
eigenvalues $\lambda_l^{(0)}$ are proportional to modified Bessel functions
$I_{l+d/2-1} (K_0)$.

The partition sum and tangent-tangent correlations may be calculated
exactly for open and periodic boundary conditions as was done, e.g., in
$d=3$  by Fisher~\cite{fisher} and Joyce~\cite{joyce}.
For arbitrary dimension $d$, 
the tangent-tangent correlation for open boundary
conditions is simply
\begin{equation}
\langle {\bf t}(0) \cdot {\bf t}(L) \rangle = 
   \left[\lambda_1^{(0)}(K_0) / \lambda_0^{(0)}(K_0) \right]^{L/b_0} \;,
\end{equation}
which reduces in the continuum limit of small $b_0$ or 
 large $K_0$ to (\ref{tangent}).

\section{Renormalization Procedure}

The real-space functional 
RG analysis for the semiflexible chain  (\ref{discrete}) 
proceeds in close analogy to the one-dimensional Heisenberg 
model \cite{niemeijer}
and  similarly to the Ising-like 
case where the ${\bf t}_i$'s are confined to discrete values
\cite{nauenberg}.  
Similar real-space functional RG methods have also been used to 
study wetting transitions or the unbinding transitions 
of strings \cite{JLM90,S91}.
In each RG step, every second tangent degree of freedom is eliminated.  
We introduce a general transfer matrix
\begin{equation}
T_{i,i-1}= \exp\left[h({\bf t}_i\cdot {\bf t}_{i-1}, K)\right]
\end{equation}
where $h=h(u,K)$ defines an {\em arbitrary} 
interaction function depending on the
scalar product of adjacent tangents $u={\bf t}_i\cdot {\bf t}_{i-1}$ 
and the  parameter $K$. 
We start the RG procedure with an initial value
$K=K_0$ and an initial interaction function $h(u, K) = -K(1-u)$, 
see eq.\ (\ref{Tij}). 
Also for an arbitrary interaction function $h(u,K)$ we can expand the
transfer matrix in the same sets of functions as in
(\ref{expansions}), which defines 
eigenvalues $\lambda_m=\lambda_m(K)$ in 2d and
$\lambda_l=\lambda_l(K)$ 
in 3d. Initially, 
these eigenvalues are given by  $\lambda_m(K)=\lambda_m^{(0)}(K)$ and
$\lambda_m(K)=\lambda_l^{(0)}(K)$, see (\ref{expansions}).

Integration over one intermediate tangent ${\bf t}'$ between 
${\bf t}$ and ${\bf t}''$ defines a recursion formula resulting
in a new interaction function $h'=h'(u,K)$ and an energy shift $g'$ by
\begin{equation} 
\exp[h'({\bf t}\cdot {\bf t}'', K)+g'(K)]
   =
\int d {\bf t}' \,\exp[\, h({\bf t} \cdot {\bf t}' ,K) 
   + h({\bf t}' \cdot {\bf t}'', K)]
,
\label{recursion1}
\end{equation}
where the energy shift $g'$ is determined by the condition that 
$h'(1, K) = h(1, K)=0$, i.e., the energy is shifted in such a way that 
the interaction term is zero for a straight polymer. 
This leads to 
\begin{equation}
\exp[g'(K)] = 
  \int d {\bf t} \, \exp[2\,h({\bf t} \cdot {\bf t}' ,K)]
.
\label{recursion2}
\end{equation}
The recursions (\ref{recursion1}) and (\ref{recursion2}) are exact 
and can be used to obtain an exact RG relation for the
eigenvalues $\lambda_k^{(N)}$ after $N$ RG recursions, 
\begin{equation}
\label{recursion_lambda}
\lambda_k^{(N+1)}= \left[\lambda_k^{(N)}\right]^2\Big/
    \left\{{\sideset{}{_{n}^{(d)}}\sum  \left[\lambda_n^{(N)}\right]^2}\right\}
 = {\left[\lambda_k^{(0)}\right]^{2^N}}\Big/
   \left\{{\sideset{}{_{n}^{(d)}}\sum 
   \left[\lambda_n^{(0)}\right]^{2^N}}\right\}
\end{equation}

In general, the new and old interactions $h'(u,K)$ and $h(u,K)$ 
will differ in their functional structure. 
Thus the renormalization of the parameter
$K$  cannot be carried out in an exact and simple manner as for 
one-dimensional Ising-like models with discrete spin orientation 
\cite{nauenberg}.  
The only fixed point function 
of the recursion (\ref{recursion1})  is  independent of $u$, i.e., 
$h^*(u,K)=0$ because of $h^*(1,K)=0$. 
This result, together with  the condition $h'(1,K)=0$,  which is
imposed  at every RG step, suggests
that  the function $h'(u,K)$ can be approximated by a
 {\em linear} function
\begin{equation}
h'(u,K) \simeq - K'(K) \left( 1-u \right)
~~~ \mbox{for}~ 
  u={\bf t} \cdot {\bf t}'' \simeq 1
,
\label{approximation}
\end{equation}
 as long as the scalar product 
$u={\bf t} \cdot {\bf t}''$
is close to one, i.e., sufficiently close to the straight 
configuration. 
This approximation should improve when the whole function 
$h'(u,K)$ becomes small upon approaching the fixed point 
$h^*(u,K)=0$ after many iterations, i.e., on large length scales. 
Using the approximation (\ref{approximation}), 
 $K'(K)$ is defined by the slope of $h'(u,K)$ at $u=1$, 
\begin{equation}
K'(K) \equiv \frac{d h'(u,K)}{d u}\Big|_{u=1} =
\frac{d}{du} \exp\big[h'(u,K)\big] \Big|_{u=1}\;,
~~~\mbox{with}~ u = {\bf t} \cdot {\bf t}''
.
\label{recursionK}
\end{equation}
Equivalently, one could expand the explicit expression for $h'(x,K)$
given by (\ref{recursion1}) and the right hand side
of  (\ref{approximation}) for small tangent angles and compare the
coefficients.
In order  to extract the renormalized bending rigidity 
$\kappa'$ from the result
for $K'$, one has to take into account that $K'$ also contains the
new bond length $b'= 2 b$, which increases by a factor of $2$ 
at each decimation step. Therefore, 
\begin{equation}
  \kappa'(K) = 2 b T \,K'(K)
.
\label{recursionkappa}
\end{equation}
Using this procedure 
we can calculate the renormalized bending rigidity $\kappa_N$ after 
$N$ RG recursions in 2d and 3d starting from the exact RG recursions 
(\ref{recursion_lambda}) for the eigenvalues. 
Inserting the renormalized eigenvalues into (\ref{expansions}), 
taking the derivative according to (\ref{recursionK}) and 
applying the rescaling (\ref{recursionkappa}) finally  yields
the result
\begin{equation}
\label{k_n}
\frac{\kappa_N}{\kappa_0}  = \frac{2^N}{K_0}
\bigg\{ \sideset{}{_{n}^{(d)}}\sum  
        \left[\lambda_n^{(0)}(K_0)\right]^{2^N} A^{(d)}_n\bigg\}
\bigg/
 \bigg\{ \sideset{}{_{n}^{(d)}}\sum  
     \left[\lambda_n^{(0)}(K_0)\right]^{2^N}\bigg\}\;,
\end{equation}
with $A^{(2)}_n\equiv n^2$ and $A^{(3)}_n\equiv\frac{1}{2}n(n+1)$.
In the following we will interpret $\kappa_N$ 
as a continuous function $\kappa(\ell)$ 
 of the length scale $\ell$ by  replacing  the rescaling factor 
 $2^N = b_N /b_0$ by the  continuous parameter $\ell/b_0$.

\begin{figure}
   \begin{center}
   \epsfig{file=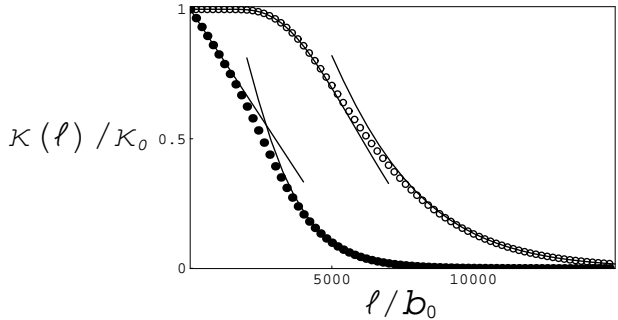,width=0.58\textwidth}
   \caption{ \label{fig:kappa_n}
  Renormalized  bending rigidity $\kappa(\ell)/\kappa_0$ as a function of 
the length scale $\ell/b_0= 2^N$ for $K_0=1000$ 
  for $d=2$ ($\circ$) and $d=3$ ($\bullet$)
  according to the recursion relation 
 (\ref{k_n}). The lines show the asymptotic 
  behavior for $\ell \gg {\kappa_0}/{T}$ and 
  $\ell \ll {\kappa_0}/{T}$ according to eqs.\  (\ref{expansion}) and 
 (\ref{expansion1}), respectively.
  }
   \end{center}
\end{figure}

\section{Persistence length}

The sums in the expressions for the effective bending
rigidity (\ref{k_n}) can be computed numerically.
Fig.\ \ref{fig:kappa_n} displays the results for 
$\kappa(\ell)/\kappa_0$ as a function of 
$\ell/b_0$ for $K_0=1000$ and in 2d and 3d.
The value 
$K_0=1000$ is appropriate for a semiflexible polymer with $\kappa_0/T
= 10 \, \mu{\rm m}$ and a bond length $b_0 = 10\, {\rm nm}$, which is
close to experimental values for F-actin \cite{gittes,kaes}.
For DNA, appropriate values are $\kappa_0/T \simeq 50 \, {\rm nm}$ 
and  $b_0 \simeq 0.3\, {\rm nm}$ and, thus, $K_0 \simeq 150$. 

As long as $\ell$ is small,
$\kappa$ decays almost linearly in $d=3$, which is also  in qualitative
agreement with the result (\ref{sigma_model}) from  the RG of the 
nonlinear $\sigma$-model. 
For $d=2$ the decay is much slower at small length scales, 
but, in contrast to the non-linear $\sigma$-model where $\kappa$ is
{\em not} renormalized. This qualitative difference is due  
to the following important 
difference between the Heisenberg and the nonlinear 
$\sigma$-model: Parametrizing the WLC model (\ref{continuous}) via 
tangent angle leaves only  quadratic terms
$\propto (\Delta\theta_{i,i-1})^2$, 
whereas the discrete 
semiflexible chain (\ref{discrete}) gives terms 
$\propto 1-\cos(\Delta\theta_{i,i-1})$,
which represent the full expansion of the cosine and  obey the local
invariance under full rotations. 

As $\ell$ increases $\kappa(\ell)$ approaches zero only asymptotically.
Therefore,  the definition of the persistence length 
as length scale where the renormalized $\kappa$ vanishes, 
$\kappa(L_p)=0$ -- which is usually used for fluid membranes --
would always give an {\em infinite} result. 
We propose  not to ask at  which length scale the renormalized
$\kappa$ reaches zero, but rather {\em how} it reaches zero. 
For $\ell \ge b_0 K_0 = \kappa_0/T$ the sums 
in (\ref{k_n}) converge  fast and one has to include only
the first few terms for  accurate results.
In fact, one can replace the Bessel functions contained in the
eigenvalues $\lambda_k^{(0)}$, see (\ref{expansions}), 
by their asymptotic form 
$I_k(x) \approx ({x}/{2 \pi})^{-1/2}
  \exp[x -(k^2 - {1}/{4})(2x)^{-1}]$ for large $x$
  \cite{AS}. 
This is justified for sufficiently large 
$K_0 \gtrsim 100$, which is fulfilled by semiflexible polymers like 
F-actin ($K_0\simeq 1000$) or DNA ($K_0\simeq 150$).
Using this asymptotics we find $(\lambda_m^{(0)} (K_0))^{\ell/b_0} \sim
e^{- {m^2 \ell}/{2 b_0 K_0}}$ for $d=2$ and 
$(\lambda_l^{(0)} (K_0))^{\ell/b_0} \sim
e^{- {l(l+1) \ell}/{2 b_0 K_0}}$ for $d=3$.
Moreover,  we may expand (\ref{k_n}) as a power series in 
$e^{-\ell T/\kappa_0}$
and obtain
\begin{equation}
\begin{split}\label{expansion}
&{\kappa (\ell)}/{\kappa_0} \approx 
 ({\ell T}/{\kappa_0})
\left( 2  e^{-\ell T / 2\kappa_0} - 4e^{-\ell T / \kappa_0} 
         + 8 e^{-3 \ell T / 2\kappa_0} -\ldots\right) ~~ \text{for}~~ d=2,\\
&
{\kappa (\ell)}/{\kappa_0} \approx 
 ({\ell T}/{\kappa_0})
\left( 3  e^{-\ell T/\kappa_0} - 9e^{-2 \ell T/\kappa_0} 
      + 42 e^{-3 \ell T/\kappa_0} -\ldots\right) ~~ \text{for}~~ d=3\;.
\end{split}
\end{equation}
The characteristic length scales in the expansions are
$2 \kappa_0/{T}$ in $d=2$ and ${\kappa_0}/{T}$ in $d=3$, which are,
therefore,  a natural definition for the persistence length $L_p$.
For general dimensionality $d$, the exponent of the first term is
determined by the order of the Bessel function appearing in
the eigenvalue.  Thus the RG calculation leads to a persistence 
length
\begin{equation}
\label{persistence_heisenberg}
L_p=\frac{2 \kappa_0}{T (d-1)}
,
\end{equation}
which agrees
 exactly with the result (\ref{tangent}) based on the tangent
 correlation function. 

Our definition based on the large-scale asymptotics of the exact RG flow
is qualitatively  different from the definition 
(\ref{persistence_sigma}) used in
perturbative RG calculations. While the result (\ref{sigma_model}) 
from the nonlinear $\sigma$-model is 
only valid for small length scales $\ell \ll {\kappa_0}/{T}$, 
where $\kappa(\ell)$ is close to $\kappa_0$, the
expansion (\ref{expansion}) describes the region $\ell \gg
{\kappa_0}/{T}$.  
Indeed, taking the expansion of (\ref{k_n}) for $\ell \ll {\kappa_0}/{T}$,
that is
\begin{equation}
\begin{split}\label{expansion1}
&
 {\kappa (\ell)}/{\kappa_0} \approx 
  1- ({8 \pi^2 \kappa_0}/{\ell T}) e^{-2 \pi^2 \kappa_0 / \ell T} + 
{\cal O}( \ell e^{-4 \pi^2 \kappa_0 / \ell T}) ~~~ \text{for} ~~ d=2, \\
&
 {\kappa (\ell)}/{\kappa_0} \approx
 1- ({\ell T}/{6 \kappa_0}) - {\cal O}(\ell^2 T^2 / \kappa_0^2) 
  \qquad\qquad\qquad\qquad  \text{for} ~~ d=3\;,
\end{split}
\end{equation}
and defining $L_p$ by the exponential decay length in two 
dimensions, respectively, 
by the linear term in three dimensions leads to a persistence length, which is
considerably bigger than the value 
(\ref{persistence_heisenberg}) found above. 
Hence it is not surprising that the two computations (\ref{sigma_model}) 
and (\ref{expansion}) yield different results.
The slow exponential decay in the expansion (\ref{expansion1}) for 
$d=2$ is reminiscent of the 
non-renormalization of $\kappa$ in the non-linear $\sigma$-model, 
see eq.\ (\ref{sigma_model}), and leads to a ``plateau''
in the numerical result for 
$\kappa(\ell)/\kappa_0$ for $\ell \ll {\kappa_0}/{T}$ in 
  Fig.\ \ref{fig:kappa_n}.

\section{Conclusion}

In conclusion we have presented a 
definition of the persistence length $L_p$ 
of a semiflexible polymer or one-dimensional fluid membrane 
based on the large scale behaviour of the 
 RG flow of the bending rigidity $\kappa$, as obtained from a
functional real-space RG calculation. 
Our result (\ref{persistence_heisenberg}) for $L_p$ 
generalizes  the conventional definition based on  the exponential
decay of a particular two-point tangent correlation function
and  gives identical results for $L_p$,  thus justifying
past experimental and theoretical work based on this conventional definition.
The  RG flows (\ref{k_n}) or (\ref{expansion}) 
allow  us to follow the behaviour 
of a semiflexible polymer from a stiff polymer on short length scales 
to an effectively flexible polymer  on large length scales 
quantitatively as a function of the length scale. 
On large length scales, our  functional RG 
gives qualitatively different results from perturbative RG 
techniques, which have been used for the closely related problem 
of  fluid membranes \cite{helfrich,
peliti,foerster,kleinert,helfrich98,pinnowhelfrich} 
and which we also applied 
to the one-dimensional semiflexible polymer. 
The generalization of our  
renormalization approach to two-dimensional  fluid membranes 
is complicated by the more involved differential
geometry of these two-dimensional objects and remains an open issue 
for future investigation.  
In analogy with our results for one-dimensional polymers and filaments
as described here, we expect that a proper definition of 
the persistence length for two-dimensional fluid membranes 
again requires the aymptotic RG flow on
large length scales which remains to be determined.


\end{document}